\documentclass[a4paper,english]{paper} 
\usepackage{babel}  
\usepackage[margin=3cm]{geometry}
\usepackage{graphicx}
\usepackage{lipsum}
\usepackage{xcolor}
\usepackage{booktabs}

\usepackage{authblk}
\usepackage{graphicx}
\usepackage{amsmath,amssymb}

\usepackage{cite} 

\title{Post-compression of picosecond pulses into the few-cycle regime}

\begin{document}
\renewcommand{\abstractname}{\vspace{-\baselineskip}} 

\author[1,2,+,*]{Prannay Balla}
\author[1,+]{Ammar Bin Wahid}
\author[3]{Ivan Sytcevich}
\author[3]{Chen Guo}
\author[3]{Anne-Lise Viotti}
\author[1,4]{Laura Silletti}
\author[5]{Andrea Cartella}
\author[1]{Skirmantas Alisauskas}
\author[1]{Hamed Tavakol}
\author[1]{Uwe Grosse-Wortmann}
\author[1,2]{Arthur Sch\"onberg}
\author[1]{Marcus Seidel}
\author[1,4]{Andrea Trabattoni}
\author[1]{Bastian Manschwetus}
\author[1]{Tino Lang}
\author[1,4,5,6]{Francesca Calegari}
\author[7]{Arnaud Couairon}
\author[3]{Anne L'Huillier}
\author[3]{Cord L. Arnold}
\author[1]{Ingmar Hartl}
\author[1,2,*]{Christoph M. Heyl}

\affil[1]{Deutsches Elektronen-Synchrotron DESY, Notkestra{\ss}e 85, 22607 Hamburg, Germany}
\affil[2]{Helmholtz-Institute Jena, Fr\"obelstieg 3, 07743 Jena, Germany}
\affil[3]{Department of Physics, Lund University, P.O Box 118, SE-221 00 Lund, Sweden}
\affil[4]{Center for Free-Electron Laser Science (CFEL), DESY, Notkestra{\ss}e 85, Hamburg, 22607, Germany}
\affil[5]{The Hamburg Centre for Ultrafast Imaging, Universit\"at Hamburg, 149 Luruper Chaussee, Hamburg 22761, Germany}
\affil[6]{Institut f\"ur Experimentalphysik, Universit\"at Hamburg, Luruper Chaussee 149, 22761 Hamburg, Germany}
\affil[7]{Centre de Physique Th\'eorique, CNRS, Ecole Polytechnique, Institut Polytechnique de Paris,  91128 Palaiseau, France}

\affil[*]{Corresponding authors: prannay.balla@desy.de; christoph.heyl@desy.de}




\maketitle
\vspace{1cm}
\begin{abstract}
In this work, we demonstrate post-compression of 1.2 picosecond laser pulses to 13\,fs via gas-based multi-pass spectral broadening. Our results yield a single-stage compression factor of about 40 at 200 W in-burst average power and a total compression factor $>$90 at reduced power. The employed scheme represents a route towards compact few-cycle sources driven by industrial-grade Yb:YAG lasers at high average power. 
\end{abstract}
\vspace{1cm}

Few-cycle laser pulses are nowadays employed for a variety of applications including attosecond physics \cite{Calegari2016jopbamaop}, light-wave electronics \cite{Sommer2016n} as well as particle acceleration \cite{Faure2018ppacf}. Their generation relies either on complex laser systems employing parametric amplification or post-compression.
Ti:Sapphire (Ti:Sa) lasers have been the working horses of ultrafast laboratories for many years. Whereas Ti:Sa amplifiers readily provide sub-30\,fs pulses, their average power is limited to only a few watts. In contrast, ultrafast Yb-based amplifiers are  power-scalable into the kW regime \cite{Nubbemeyer2017ol,Mueller2018ol}, however, their output pulse duration is not shorter than 100 fs or even 1 ps. External spectral broadening and pulse-post compression offer the possibility to combine ultrashort pulse durations and high repetition rates with mJ-level pulse energies \cite{Nagy2019o,Fan2016o}. 
Pulses at the few-mJ level are routinely post-compressed via spectral broadening in gas-filled hollow-core fibers (HCFs) \cite{Bohle2014lpl,Fan2016o}. While HCF-based post-compression setups extend to large scales at multi-mJ pulse energies and/or high compression ratios \cite{Jeong2018sr,Nagy2019o}, a more recent scheme, multi-pass cell (MPC) -based post-compression \cite{Schulte2016ol,Hanna2017jotosoab}, allows compact setups and compression factors $>$30 \cite{Kaumanns2018ol}.
Especially when using a gas as nonlinear medium \cite{Lavenu2018ol,Ueffing2018ol}, MPC-based post-compression offers great prospects for peak-power scaling \cite{Heyl2016o,Kaumanns2018ol,Russbueldt2019ol}; the gas filling is immune to damage and the nonlinearity is easily adapted via the pressure.
However, in contrast to hollow-core fiber compression setups, which are commonly employed to reach the few-cycle regime \cite{Bohle2014lpl,Louisy2015o,Fan2016o,Jarque2018sr}, MPC-based pulse-compression yielded typically sub-100 \cite{Schulte2016ol,Hanna2017jotosoab,Weitenberg2017oe,Weitenberg2017ijoqe,Lavenu2018ol,Ueffing2018ol,Kaumanns2018ol,Russbueldt2019ol,Jargot2018ol,Tsai2019ol,Vicentini2020oe} down to sub-20\,fs pulses \cite{Fritsch2018ola}. 
Broader spectra supporting few- to single-cycle pulses have been generated by multi-plate arrangements \cite{Lu2014o,Lu2019oe,Seo2020ol}. The approach is similar to the MPC method. However, the refocusing of the beam is accomplished by the nonlinear media themselves. The drawbacks of the multi-plate approach are the practically limited number of passes and the required higher nonlinear phase accumulation per pass which results in imperfect spectral homogeneity of the beam \cite{Milosevic2000ol,Seidel2016oe}. 

We here extend MPC-based post-compression into the few-cycle regime via a dual-stage setup, driven with 1.2\,ps pulses. 
In a first compression stage, we spectrally broaden and compress 2 mJ pulses at 200 W in-burst average power to 32 fs, followed by further compression to 13 fs at reduced pulse energy in the second stage.
These results demonstrate, to the best of our knowledge, for the first time few-cycle pulse generation via direct post-compression of picosecond pulses, opening a route towards high-average power few-cycle sources driven by kW-class industrial-grade diode pumped Yb:YAG lasers.
In addition, we push the single-stage compression ratio achieved via mJ-level post-compression close to 40 with $>$80\% throughput, surpassing recent compression factors achieved using stretched hollow-core fibers \cite{Jeong2018sr, Nagy2019o} and MPCs \cite{Kaumanns2018ol}.  

%
\begin{figure*}[htb!]
\centering
\includegraphics[width=.9\linewidth]{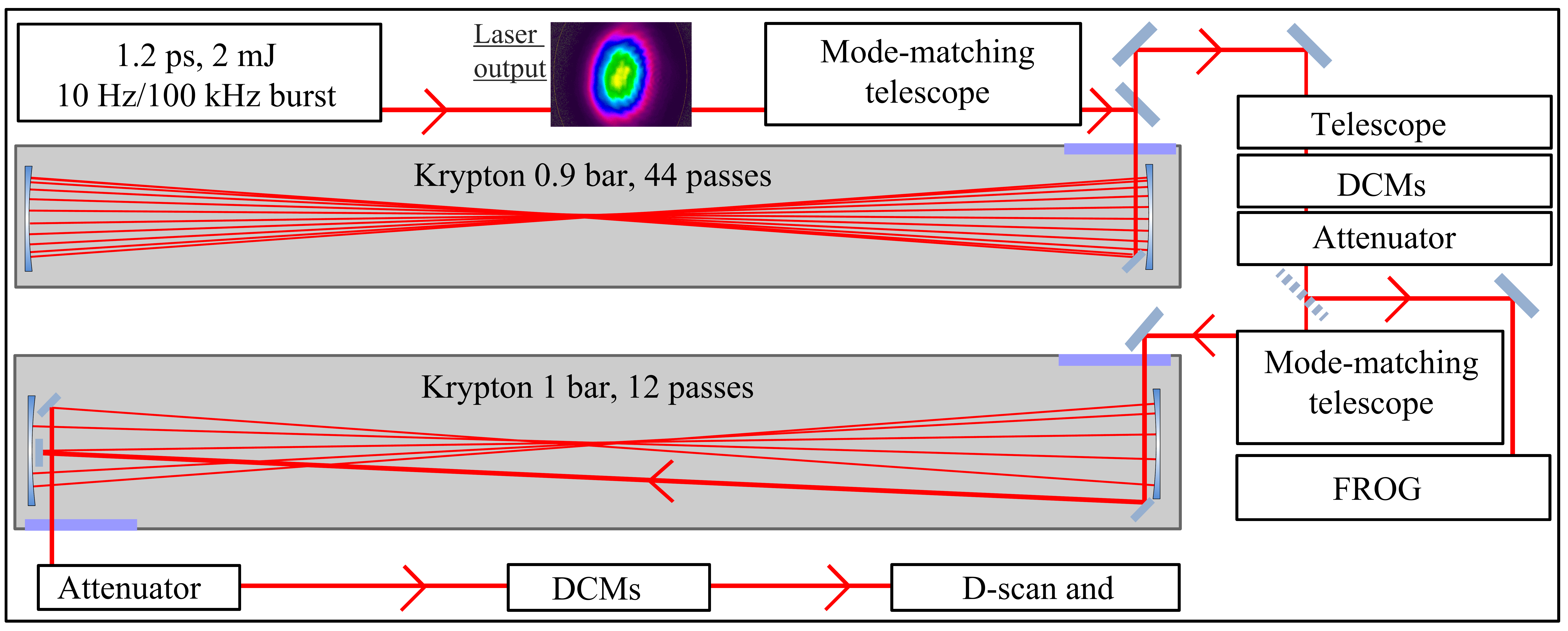}
\caption{Experimental setup: Two cascaded gas-filled multi-pass cells (length~$\lesssim$\,2~m) followed by DCM compressors are used for spectral broadening and compression.}
\label{fig:setup}
\end{figure*}
The experimental setup is depicted in Figure~\ref{fig:setup}. It consists of two broadening and compression stages containing  gas-filled MPCs, mode-matching units and dispersion compensating mirrors (DCMs). The employed Innoslab laser amplifier provides 1.2\,ps (FWHM) pulses centered at 1030 nm at an M$^2$ of 1.1x1.2, operated at a pulse energy of 2\,mJ at the entrance of the first spectral broadening unit \cite{Lang20192colaeeeqecc}. The laser runs in burst-mode, i.e. pulse-trains of about 100 pulses at 100 kHz intra-burst repetition rate are emitted at 10 Hz. The pulses are coupled to a first MPC, which consists of two dielectric concave mirrors (1~m radius of curvature, 100~mm diameter, GDD <10~fs$^2$ within a 980-1080~nm bandwidth), set-up in a vacuum chamber flooded with krypton at 0.9~bar. The laser beam is mode-matched to the MPC eigenmode using a mode-matching telescope. 
\begin{figure}[bht]
\centering
\includegraphics[width=.5\linewidth]{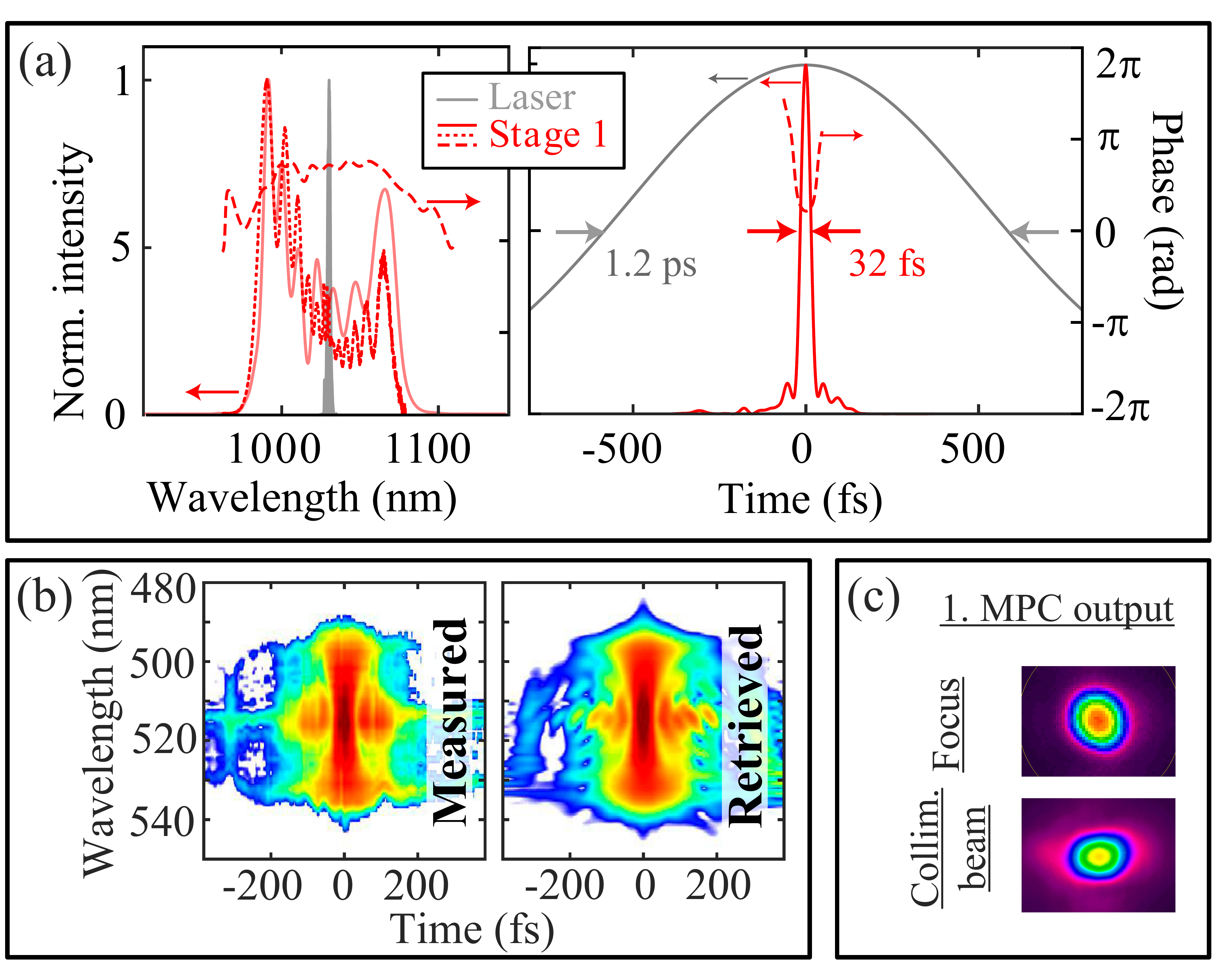}
\caption{(a) Reconstructed spectral and temporal intensity (solid) and phase (dashed) profiles together with the corresponding spectra measured after the first MPC (dotted) and at the laser output (gray line/area). (b) Corresponding FROG traces (logarithmic color scale). (c) Beam profiles measured after the first MPC.}
\label{fig:results1}
\end{figure}

In- and out-coupling is realized through an anti-reflection coated window and a rectangular pick-off mirror placed in front of one of the concave mirrors. 
The same mirror is also used to couple the beam out after 44 passes though the MPC. The pulses are compressed using 32 bounces on DCMs with a total GDD of -6200\,fs$^2$ in a bandwidth of 980-1080\,nm. 
The throughput of the first MPC is 85\% (80\% including the DCM compressor), yielding compressed pulses at 1.6 mJ and 160 W in-burst average power. The broadened spectrum has a -10~dB bandwidth of 90~nm, corresponding to a Fourier-limited pulse duration (FWHM) of 30~fs. The pulses are compressed to 32~fs, measured via second harmonic frequency resolved optical gating (Figure~\ref{fig:results1}(a,b)). The beam profiles of the collimated and focused beam measured after the DCM compressor (c) indicate good beam quality.

\begin{figure}[htb]
\centering
\includegraphics[width=.6\linewidth]{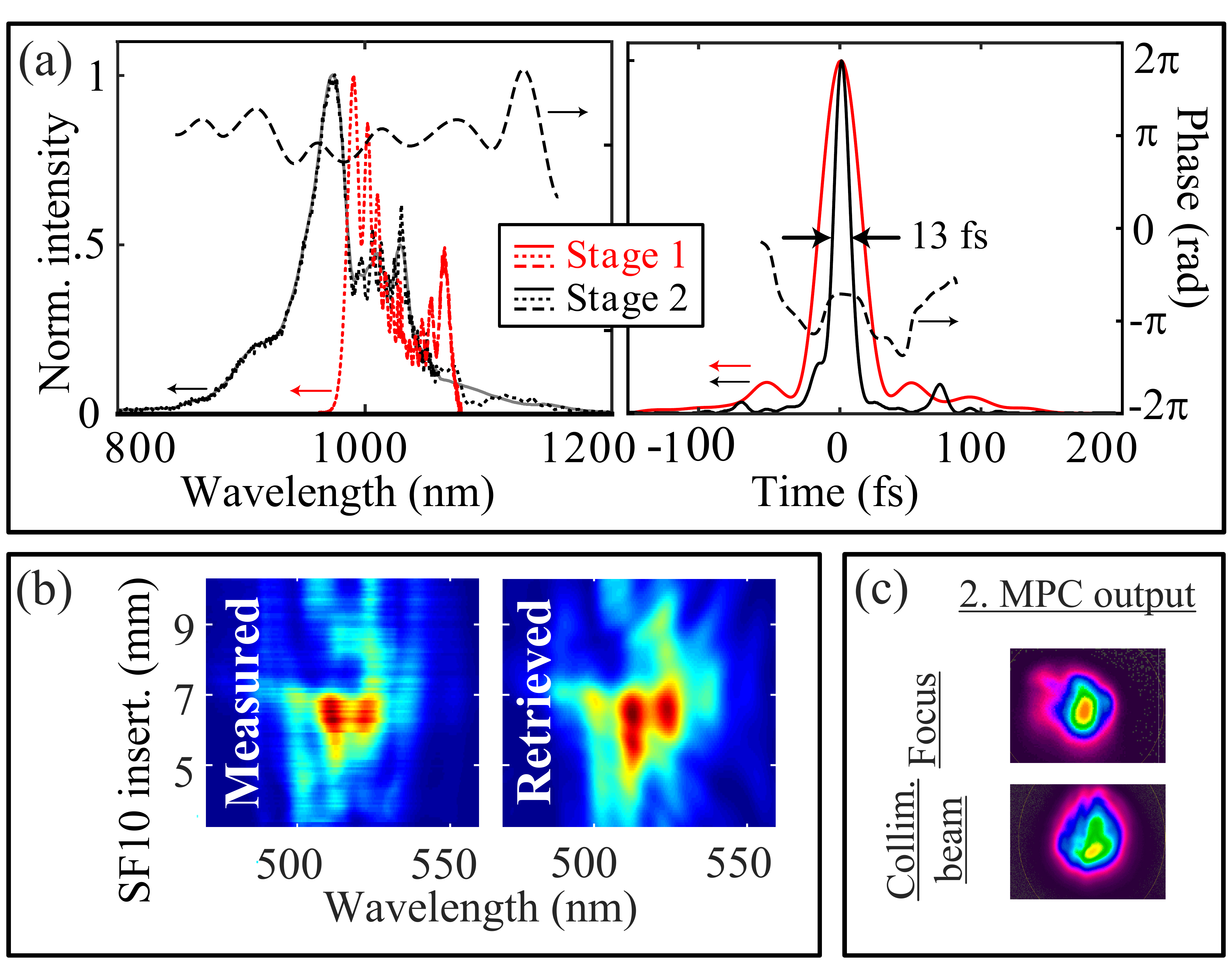}
\caption{(a) Reconstructed spectral and temporal intensity (solid) and phase (dashed) profiles together with the corresponding spectra measured after both MPCs (dotted). (b) Corresponding D-scan traces (linear color scale). (c) Beam profiles measured after the second MPC.}
\label{fig:results2}
\end{figure}

To further compress the pulses while mitigating bandwidth limitation imposed by the mirror coating of the first MPC, we send the laser beam to a second MPC consisting of two silver mirrors with enhanced reflectivity (1~m radius of curvature, 75~mm diameter). In order to avoid MPC mirror damage, the laser pulses are attenuated to about 0.8~mJ. The second MPC is filled with krypton at a pressure of 1~bar.
The laser beam enters the second MPC chamber via an anti-reflection coated window and leaves through an uncoated window. In order to reduce the fluence at the in-coupling window, the converging input beam is propagated about 2\,m through the vacuum chamber before in-coupling into the MPC. 
Separate mirrors are used for in-coupling and out-coupling after 12 passes.
The measured throughput of the second MPC is 46\%, limited mainly by losses at the MPC mirrors, yielding a pulse energy of 0.37~mJ at an in-burst average power of 37~W after the out-coupling window.
To avoid nonlinear propagation effects in air, only a wedge reflection of the pulses exiting the MPC is compressed using matched-pair DCMs with a total GDD of -1900~fs$^2$ in 16 double bounces. The measured output spectrum has a bandwidth of 195~nm at -10~dB, corresponding to a Fourier limit of 10.8~fs. The attenuated pulses are characterized using a D-scan setup yielding a pulse duration (FWHM) of 13~fs (Figure~\ref{fig:results2} (a,b)). 
While we achieve good agreement between measured and retrieved D-scan traces (b), the measured traces clearly show intensity fluctuation of the second harmonic signal, which we attribute to laser pulse duration fluctuations.
The beam profiles of the collimated and focused beam measured after the second MPC (Figure~\ref{fig:results2} (c)) indicate a reduced beam quality compared to the first cell output. A possible reason for this could be heating of the silver mirrors caused by absorption or unwanted nonlinear propagation effects, which could arise in the attenuator before the second MPC. 

Because of the great power-scalability offered by gas-filled MPCs, the compression of laser pulses with higher pulse energies seems feasible in the first compression stage of our setup. While the nonlinearity could easily be adjusted by a decreased gas density at increased pulse energy \cite{Kaumanns2018ol}, limitations are expected due to laser induced damage. In our experimental configuration, the use of 2~mJ pulses corresponds to a fluence of 30\,mJ/cm$^2$ at the mirrors of the first MPC. 
Limitation caused by ionization, which are prevented in our experiment by an appropriate choice of the MPC geometry, may arise at higher pulse energies. However, they can easily be circumvented by employing a gas with higher ionization potential and/or by increasing the MPC size.

The overall transmission through our setup was severely limited by the employed mirrors in the second MPC. While silver mirrors typically provide a flat GDD over a large bandwidth, their reflectivity limits the transmission. In addition, in our experiment, observed laser induced damage of the silver mirrors at full pulse energy (1.6~mJ) demanded the operation of the second MPC at reduced power. The use of high-damage threshold multi-layer mirrors in the second MPC could circumvent this limitation. Also, optimization of the broadening process in the second MPC could be realized via dispersion engineering e.g. to mitigate temporal pulse broadening induced by linear dispersion (approximately 800\,fs$^2$ in the second MPC). Taking into account the compressor parameters and applying them to the measured pulse, we estimate a temporal pulse broadening to about 200\,fs in the second MPC. Such a severe temporal broadening effectively restricts spectral broadening to the first few passes through the MPC. This effect might also contribute to a reduced spatial beam quality.     

In conclusion, we have demonstrated post-compression of picosecond laser pulses into the few-cycle regime yielding a reduction in pulse duration by about two orders of magnitude. Moreover, we achieve a record single-stage compression factor of about 40 at the mJ level employing a 2.5 \,m long setup. Our proof-of-principle demonstration was limited mainly by damage threshold, reflectivity and the lack of dispersion control imposed by the silver mirrors employed in the second compression stage.  
We expect that the use of dispersion engineered, high damage threshold, dielectric mirrors in a second broadening stage together with gas-based nonlinear spectral broadening will unfold further pulse duration down-scaling and pulse energy up-scaling options. 
As the employed laser and compression scheme is known for compactness as well as excellent pulse energy and average power scalability \cite{Schmidt2017oe,Kaumanns2018ol}, our methods promises further pulse energy upscaling potential and may thus enable future TW-class, few-cycle laser sources driven by industrial-grade kW-scale picosecond lasers. Such laser sources will not only open completely new parameter regimes for attosecond physics and strong-field science, they also provide a viable route for next generation laser-driven electron accelerators.

\section*{Funding Information}
This work is supported by a PIER Seed Projects grant (partnership of Universit\"at Hamburg and DESY, PIF-2018-85), the Cluster of Excellence 'CUI: Advanced Imaging of Matter' of the Deutsche Forschungsgemeinschaft - EXC 2056 (390715994), the Swedish Research Council and the European Research Council (339253 PALP). 

\section*{Disclosures}
The authors declare no conflicts of interest.
\vspace{4mm}\\
\noindent + These authors contributed equally to this work

%



\end{document}